\documentclass[prapplied, reprint, superscriptaddress, floatfix, longbibliography]{revtex4-2}

\usepackage{mathtools}
\usepackage{hyperref}
\usepackage{braket}
\usepackage{multirow}
\usepackage{placeins}
\usepackage{mleftright}
\usepackage[norefs, nocites, ignoreunlbld]{refcheck}
\usepackage{physics}

\begin{document}

\title{Generation of Frequency-Tunable Shaped Single Microwave Photons Using a Fixed-Frequency Superconducting Qubit}

\author{Takeaki~Miyamura}
\email{miyamura@qipe.t.u-tokyo.ac.jp}
\affiliation{Department of Applied Physics, Graduate School of Engineering, The University of Tokyo, Bunkyo-ku, Tokyo 113-8656, Japan}
\author{Yoshiki~Sunada}
\altaffiliation[Present address: ]{QCD Labs, QTF Centre of Excellence, Department of Applied Physics, Aalto University, P.O.\ Box 13500, FIN-00076 Aalto, Finland}
\affiliation{Department of Applied Physics, Graduate School of Engineering, The University of Tokyo, Bunkyo-ku, Tokyo 113-8656, Japan}
\author{Zhiling~Wang}
\affiliation{RIKEN Center for Quantum Computing (RQC), Wako, Saitama 351-0198, Japan}
\author{Jesper~Ilves}
\affiliation{Department of Applied Physics, Graduate School of Engineering, The University of Tokyo, Bunkyo-ku, Tokyo 113-8656, Japan}
\author{Kohei~Matsuura}
\affiliation{Department of Applied Physics, Graduate School of Engineering, The University of Tokyo, Bunkyo-ku, Tokyo 113-8656, Japan}
\author{Yasunobu~Nakamura}
\affiliation{Department of Applied Physics, Graduate School of Engineering, The University of Tokyo, Bunkyo-ku, Tokyo 113-8656, Japan}
\affiliation{RIKEN Center for Quantum Computing (RQC), Wako, Saitama 351-0198, Japan}

\date{\today}

\begin{abstract}
Scaling up a superconducting quantum computer will likely require quantum communication between remote chips, which can be implemented using an itinerant microwave photon in a transmission line. 
To realize high-fidelity communication, it is essential to control the frequency and temporal shape of the microwave photon.
In this work, we demonstrate the generation of frequency-tunable shaped microwave photons without resorting to any frequency-tunable circuit element.
We develop a framework which treats a microwave resonator as a band-pass filter mediating the interaction between a superconducting qubit and the modes in the transmission line.
This interpretation allows us to stimulate the photon emission by an off-resonant drive signal.
We characterize how the frequency and temporal shape of the generated photon depends on the frequency and amplitude of the drive signal.
By modulating the drive amplitude and frequency, we achieve a frequency tunability of 40 MHz while maintaining the photon mode shape time-symmetric.
Through measurements of the quadrature amplitudes of the emitted photons, we demonstrate consistently high state and process fidelities around 95\% across the tunable frequency range. 
Our hardware-efficient approach eliminates the need for additional biasing lines typically required for frequency tuning, offering a simplified architecture for scalable quantum communication. 
\end{abstract}

\maketitle

\section{Introduction}
Fault-tolerant quantum computing using superconducting qubits requires a large number of physical qubits~\cite{fowler_surface_2012}. 
Integrating these qubits on a single chip or in a single cryostat poses significant technical challenges, including frequency crowding of the qubits~\cite{hertzberg_laser-annealing_2021} and heat-loading from the control wiring~\cite{krinner_engineering_2019}.
These challenges have motivated the development of distributed quantum computing architectures~\cite{cirac_quantum_1997, ang_arquin_2024, mollenhauer_high-efficiency_2024}, where quantum information is processed across multiple nodes connected via quantum communication channels.
Several schemes of quantum communication between remote superconducting qubits have been demonstrated~\cite{zhong_violating_2019, 
leung_deterministic_2019, chang_remote_2020, zhong_deterministic_2021, burkhart_error-detected_2021, niu_low-loss_2023, song_realization_2024} using techniques developed in circuit quantum electrodynamics~\cite{blais_circuit_2021}.

An itinerant microwave photon propagating in a transmission line offers a promising medium for such quantum communication, providing long-range information transfer and compatibility with superconducting circuits~\cite{kurpiers_deterministic_2018, axline_-demand_2018, campagne-ibarcq_deterministic_2018,  kurpiers_quantum_2019, magnard_microwave_2020, qiu_deterministic_2025, storz_loophole-free_2023, grebel_bidirectional_2024, almanakly_deterministic_2025}. 
In this approach, the sender and receiver qubits each need to couple to an interface mode~(a resonator or qubit), which has the same frequency as the photon.
To transfer one qubit of information, the sender qubit first encodes its information into an itinerant photon by executing a conditional photon emission~\cite{korotkov_flying_2011,pierre_storage_2014}.
The photon then propagates through the transmission line and is captured by the receiver qubit~\cite{wenner_catching_2014, lin_deterministic_2022}. 
However, two challenges must be addressed to realize a high-fidelity communication with this method. 
First, the absorption process requires a frequency matching between the transmitted photon and the receiver interface mode, which is hindered by frequency variations across devices due to fabrication uncertainties.
Second, the temporal shape of the emitted photon needs to be controllable~\cite{cirac_quantum_1997}.

To address these problems, previous approaches have relied on additional control mechanisms.
The photon frequency has been controlled through the use of flux-biased circuits~\cite{peng_tuneable_2016, zhou_tunable_2020, kurpiers_deterministic_2018, kannan_-demand_2023, li_frequency-tunable_2023} or mechanical positioning of a superconducting pin in proximity to the interface resonator~\cite{axline_-demand_2018}. 
The temporal shape of the photon has been controlled by using SQUID-based modulation of transmission-line modes~\cite{forn-diaz_-demand_2017}, implementing tunable couplers with flux-biased circuits~\cite{li_control_2022, yang_deterministic_2023, li_-demand_2025, nie_parametrically_2024, dong_shaping_2024}, or modulating the amplitude of microwave control drives~\cite{campagne-ibarcq_deterministic_2018, pechal_microwave-controlled_2014}. 
Although these approaches can realize the frequency tuning and photon shaping, they rely on additional flux lines, which introduce complexity when scaling up the system. 
These considerations motivate the development of simpler, more scalable approaches to frequency-tunable photon generation.

In this work, we demonstrate a hardware-efficient approach to generate frequency-tunable shaped microwave photons using a fixed-frequency circuit where a superconducting qubit is dispersively coupled to a resonator~[Fig.~\ref{fig:1}(a)].
Whereas conventional approaches have relied on resonant driving between two energy levels of the coupled system~[Fig.~\ref{fig:1}(b)], we use an off-resonant driving scheme based on treating the resonator as an effective band-pass filter acting on the qubit--transmission-line coupling~[Fig.~\ref{fig:1}(c)].
This allows us to tune the photon frequency by detuning the drive frequency and eliminates the need for an additional tuning mechanism. 
The temporal profile of the emitted photon can be controlled using a simplified model which adiabatically eliminates the resonator dynamics.
Through this approach, we achieve frequency tuning of a shaped emitted photon in a range of 40~MHz, which is sufficient to compensate for the typical fabrication variance of the resonator frequency of about~$10$~MHz~\cite{valles-sanclemente_post-fabrication_2023, li_optimizing_2023}. 
We measure the quadrature amplitudes of the emitted photons, demonstrating consistently high state and process fidelities around 95\% across the tunable frequency range. 
By eliminating the need for additional flux lines, our scheme enables hardware-efficient quantum communication between distributed nodes.

\section{Off-resonant generation of an itinerant microwave photon}\label{chap2_idea}
\begin{figure}[t]
    \centering
    \includegraphics{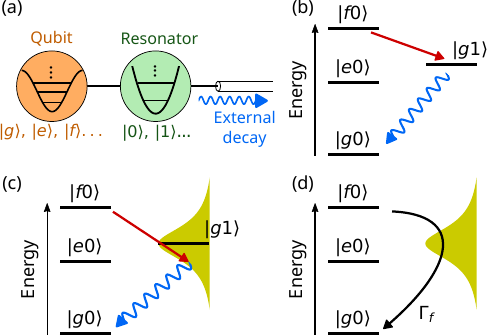}
    \caption{Emission of a frequency-tunable photon. 
    (a)~Schematic of the system.
    (b)~Conventional scheme for generating an itinerant microwave photon through a resonator-assisted Raman process activated by resonant driving of the $|f0\rangle$--$|g1\rangle$ transition. 
    (c)~Off-resonant driving~(red) within the resonator linewidth~(yellow) for frequency tuning of the emitted photon. 
    (d)~Adiabatic elimination of the resonator state. $\Gamma_f$ is defined as the effective decay rate. The yellow Lorentzian shape represents the modes on the transmission line filtered by the resonator.
    }
    \label{fig:1}
\end{figure}

We consider a system where a superconducting qubit is dispersively coupled to a resonator~[Fig.~\ref{fig:1}(a)].
The resonator is also coupled to an external transmission line.
The energy-level diagram of such a system is shown in Fig.~\ref{fig:1}(b), where $\ket{g}$,~$\ket{e}$, and~$\ket{f}$ are the ground, first excited, and second excited states of the qubit, and $\ket{0}$ and $\ket{1}$ are the photon number states of the resonator.
It has been demonstrated that an itinerant photon can be generated with this system through a resonator-assisted Raman process~\cite{pechal_microwave-controlled_2014, zeytinoglu_microwave-induced_2015}.
In this process, one applies a microwave drive to induce the coupling between $\ket{f0}$ and $\ket{g1}$, which leads to an emission of a photon from the resonator into the transmission line.
Because the strength and phase of the coupling can be controlled by the microwave drive, one can modulate the shape of the emitted photon~\cite{pechal_microwave-controlled_2014}.

Here we interpret the resonator as an effective band-pass filter acting on the qubit--transmission-line interaction. 
This interpretation reveals that the qubit effectively couples to the transmission-line modes near the resonator frequency, with the density of states shaped by the filtering effect of the resonator.
By driving at the frequency difference between the $\ket{f}$ state and these filtered modes, a photon can be emitted into the transmission line even when the drive is off-resonant from the $\ket{f0}$--$\ket{g1}$ transition~[Fig.~\ref{fig:1}(c)].
This differs from the conventional method where photons are emitted through resonant driving~[Fig.~\ref{fig:1}(b)].
In this off-resonant driving regime, the system must satisfy the energy conservation relation of $\omega_{ge} + \omega_{ef} = \omega_{\mathrm{d}} + \omega_{\mathrm{ph}}$, where $\omega_{ge}\,(\omega_{ef})$ is the transition frequency between $\ket{g}$ and $\ket{e}$ ($\ket{e}$ and $\ket{f}$) of the qubit, $\omega_{\mathrm{d}}$ is the drive frequency, and $\omega_{\mathrm{ph}}$ is the emitted-photon frequency.
Therefore, the photon frequency can be tuned by changing the detuning of the drive signal.

When emitting a photon using an off-resonant drive, it is essential to characterize the dependence of the photon emission process on the detuning.
Here, we assume that the external decay rate of the resonator, $\kappa$,  is sufficiently larger than the drive-induced coupling strength between $\ket{f0}$ and $\ket{g1}$ such that we can approximately eliminate the resonator state when modeling the system dynamics.
In this case, the drive can be seen as stimulating an energy decay of the qubit from $\ket{f}$ to $\ket{g}$.
We denote this effective decay rate induced by the drive as $\Gamma_f$~[Fig.~\ref{fig:1}(d)].
Since it determines how quickly an excitation in the qubit can be released into the transmission line, $\Gamma_f$ can be considered as the ``photon emission rate''.
In the system shown in Fig.~\ref{fig:1}(a), $\Gamma_f$ can be calculated as
\begin{equation}\label{Gamma_f_single}
    \Gamma_f(\omega_{\mathrm{ph}}) = \frac{\kappa g_{\mathrm{eff}}^2}{(\omega_{\mathrm{r}}-\omega_{\mathrm{ph}})^2 + (\kappa/2)^2},
\end{equation}
where $g_{\mathrm{eff}}$ is the drive-induced coupling strength between $\ket{f0}$ and $\ket{g1}$ and $\omega_{\mathrm{r}}$ is the resonance frequency of the resonator (see Appendix~\ref{app:photon_shaping} for the derivation).
Equation~\eqref{Gamma_f_single} shows that $\Gamma_f$ exhibits the Lorentzian spectrum of the resonator as a function of the photon frequency $\omega_{\mathrm{ph}}$.
This behavior directly reflects the resonator's role as a band-pass filter between the qubit and the transmission-line modes.
From this perspective, the measurement of $\Gamma_f$ as a function of the drive frequency allows us to probe the frequency dependence of the coupling strength between the qubit and the modes in the transmission line, which determines the frequency dependence of the photon emission process.
Moreover, the measured $\Gamma_f(\omega_{\mathrm{ph}})$ serves as a look-up table when controlling the photon emission dynamics, as $g_{\mathrm{eff}}$ and $\omega_{\mathrm{ph}}$ can be tuned through the amplitude and frequency of the drive.
In other words, one can engineer the shape of the emitted photon by modulating these parameters.
Under the adiabatic elimination condition $\kappa\gg g_{\mathrm{eff}}$, the required time dependence of the photon emission rate is
\begin{equation} \label{gamma_t} 
\Gamma_f^{\mathrm{target}}(t) = \frac{|\psi_{\mathrm{ph}}^{\mathrm{target}}(t)|^2}{1 - \int^t_{-\infty}|\psi_{\mathrm{ph}}^{\mathrm{target}}(t^\prime)|^2 \,dt^\prime},
\end{equation}
where $\psi_{\mathrm{ph}}^{\mathrm{target}}(t)$ is the target photon waveform~\cite{srinivasan_time-reversal_2014} (see Appendix~\ref{app:photon_shaping} for the derivation).

\section{Experiment}
\begin{figure}
    \centering
    \includegraphics{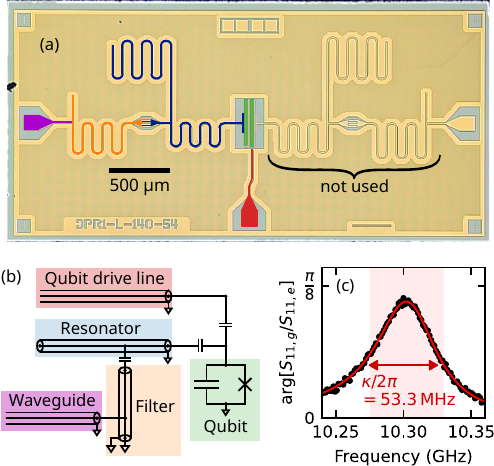}
    \caption{(a)~False-colored photograph of the device. A fixed-frequency transmon (green) is coupled to a fixed-frequency resonator (blue). The circuit incorporates an intrinsic Purcell filter in the resonator and has an additional band-pass Purcell filter~(orange) which couples to a waveguide~(purple). The qubit is driven via a dedicated drive line (red).
    (b)~Equivalent circuit diagram.
    (c)~Reflection spectroscopy of the resonator. We prepare the qubit in the $\ket{g}$~($\ket{e}$) state and measure the reflection coefficient $S_{11,g(e)}$. The black dots are the phase of the ratio $S_{11,g}/S_{11,e}$, and the red curve is the fitting result~(see Appendix~\ref{app_setup}). The fitted linewidth $\kappa$ is indicated by the arrow and shaded area.
    }
    \label{fig:2}
\end{figure}
\subsection{Device}
\begin{figure*}
    \centering
    \includegraphics{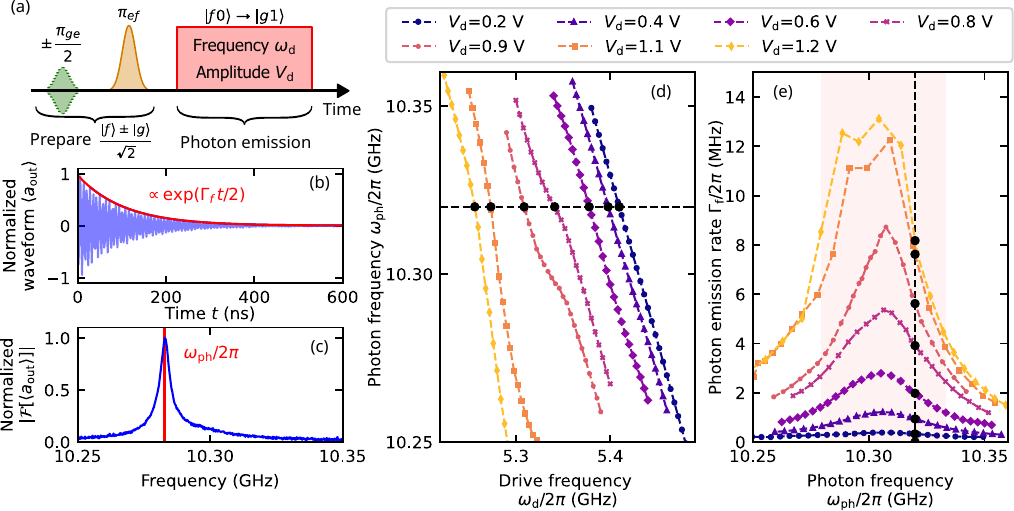}
    \caption{Characterization of the off-resonant photon-emission process. 
    (a)~Pulse sequence.
    (b)~Emitted-photon waveform for the drive amplitude $V_{\mathrm{d}}=0.8\mathrm{\,V}$ and frequency $\omega_{\mathrm{d}}/2\pi=5.384\mathrm{\,GHz}$. The fitted exponentially decaying envelope is shown as the red curve. 
    (c)~Normalized Fourier amplitude (blue) of the measured photon waveform shown in~(b). The center frequency of the photon $\omega_{\mathrm{ph}}$ is determined from the position of the peak~(red).
    (d)~Emitted-photon frequency $\omega_{\mathrm{ph}}$ as a function of $\omega_\mathrm{d}$ at various drive amplitudes $V_{\mathrm{d}}$. The values at the photon frequency of 10.32~GHz (black circles on the horizontal dashed line) are used in Sec.~\ref{chap:calibration} for designing the drive pulse. 
    (e)~Measured $\Gamma_f$ as a function of the measured photon frequency $\omega_\mathrm{ph}$ at various drive amplitudes $V_{\mathrm{d}}$. 
    The shaded region represents the resonator linewidth $\kappa$ centered around the resonance frequency $\omega_\mathrm{r}$. The values of $\Gamma_f$ at the photon frequency of 10.32~GHz (black circles on the vertical dashed line) are used in Sec.~\ref{chap:calibration}.
    }
    \label{fig:3}
\end{figure*}

The device used in this study is depicted in Fig.~\ref{fig:2}(a). 
It consists of a fixed-frequency transmon qubit~\cite{koch_charge-insensitive_2007} coupled to two resonator--filter systems, although only one of them is used.
The transmon qubit has a transition frequency of $\omega_{ge}/2\pi = 8.0259\mathrm{\,GHz}$ between the ground and first excited states and an anharmonicity of $\alpha/2\pi = -318.5\mathrm{\,MHz}$. 

For our frequency-tunable photon generation, a large resonator linewidth is desirable. 
However, this makes the qubit vulnerable to energy relaxation through the resonator, i.e., Purcell effect.
To prevent the rapid qubit decay due to the Purcell effect while maintaining a large resonator linewidth, we implement both a band-pass Purcell filter~\cite{sete_quantum_2015,jeffrey_fast_2014} and an intrinsic Purcell filter \cite{sunada_fast_2022, spring_fast_2024}, achieving an effective resonator linewidth of $\kappa/2\pi = 53.3\,\mathrm{MHz}$~[Fig.~\ref{fig:2}(c)] while maintaining a qubit energy-relaxation time of $T_1 \approx 28\,\mu\mathrm{s}$.
The resonance frequency of the resonator mode is $\omega_{\mathrm{r}}/2\pi = 10.306\mathrm{\,GHz}$ when the qubit is in the ground state.
Other device parameters are provided in Appendix~\ref{app_setup}.

\subsection{Characterizing the frequency-dependent emission rates}\label{raman_spectroscopy}
We first characterize the frequency-dependent coupling between the qubit and the transmission-line modes by measuring the photon emission rate $\Gamma_f$. 
This measurement is carried out by off-resonantly driving the $\ket{f0}$--$\ket{g1}$ transition with a square pulse and observing the waveform of the emitted photon.
The experimental protocol follows the pulse sequence shown in Fig.~\ref{fig:3}(a).
The qubit is prepared in either of the states $(\ket{g}\pm\ket{f})/\sqrt{2}$ to emit an itinerant photon in the superposition $(\ket{0}\pm\ket{1})/\sqrt{2}$.
We then take a difference of the measured waveforms with opposite signs to cancel background noise while preserving the temporal shape of the signals~\cite{sunada_efficient_2024}.
We repeat this experiment while sweeping the drive frequency~$\omega_{\mathrm{d}}$ and amplitude~$V_\mathrm{d}$, the latter of which is measured in units of volts at the arbitrary waveform generator (AWG) modulating the drive signal.
The emitted photon is amplified using a flux-driven Josephson parametric amplifier (JPA) operated in the phase-sensitive mode~\cite{yamamoto_flux-driven_2008}.
We alternate the JPA pump phase between $0$ and $\pi$ and add the results to emulate a phase-preserving amplification~\cite{sunada_efficient_2024}.

Figure~\ref{fig:3}(b) shows the measured photon waveform for the drive amplitude $V_{\mathrm{d}} = 0.8\mathrm{\,V}$ and the drive frequency $\omega_{\mathrm{d}}/2\pi = 5.384\mathrm{\,GHz}$.
Under the adiabatic elimination condition $\kappa \gg g_{\mathrm{eff}}$, the emitted photon has an exponentially decaying envelope with a rate $\Gamma_f/2$ (see Appendix~\ref{app:waveform_square_drive}), which we fit to extract $\Gamma_f$.
We also determine the central frequency of the emitted photon, $\omega_\mathrm{ph}$, by performing the Fourier transform of the measured waveform~[Fig.~\ref{fig:3}(c)].

Figure~\ref{fig:3}(d) shows the relationship between the emitted-photon frequency $\omega_{\mathrm{ph}}$ and the drive frequency $\omega_{\mathrm{d}}$.
As the drive amplitude $V_{\mathrm{d}}$ increases, the frequency relationship between $\omega_{\mathrm{d}}$ and $\omega_{\mathrm{ph}}$ shifts toward lower drive frequencies.
We attribute this fact to the ac Stark shift of the qubit states due to the $\ket{f0}$--$\ket{g1}$ drive.
We also observe that this frequency relationship becomes nonlinear at certain drive amplitudes. 
While the exact origins of this non-linear behavior remain unclear, they may be partially due to the inherent non-linear dependence of the emitted-photon frequency on the drive parameters (see Appendix~\ref{app_waveform}).
Additionally, the $\lambda/2$ resonator has an anti-resonance near the $\ket{f0}$--$\ket{g1}$ transition frequency, potentially resulting in a frequency-dependent drive strength and the non-linear frequency relationship at specific drive amplitudes.
Regardless of the underlying mechanisms, the results demonstrate that the photon frequency can be tuned by changing the drive frequency.

Figure~\ref{fig:3}(e) shows the extracted photon emission rate $\Gamma_f$ as a function of the emitted-photon frequency $\omega_\mathrm{ph}$.
The rate exhibits a Lorentzian-like dependence centered at the resonator frequency $\omega_{\mathrm{r}}$ with a width corresponding to the resonator linewidth $\kappa$ observed in Fig.~\ref{fig:2}(c). 
As we will demonstrate in the next subsection, these spectroscopic results enable us to design appropriate drive pulses for generating photons with desired shapes and frequencies.

\subsection{Designing the drive pulse for photon shaping}\label{chap:calibration}
\begin{figure}[tb]
    \centering
    \includegraphics{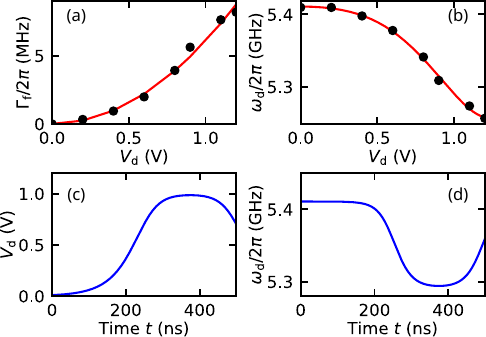}
    \caption{Designing the drive pulse for the $\ket{f0}$--$\ket{g1}$ transition to emit a photon with the desired shape $\psi_{\mathrm{target}}(t)$ and the desired frequency of 10.32~GHz.
    (a)~Photon emission rate $\Gamma_f$ as a function of the drive amplitude $V_{\mathrm{d}}$ at the target photon frequency. Black circles are the values obtained from the intersections in Fig.~\ref{fig:3}(e). Red curve shows the fitting with an even-polynomial model.
    (b)~Drive frequency $\omega_\mathrm{d}$ as a function of the drive amplitude $V_{\mathrm{d}}$. The black circles are obtained from the intersections in Fig.~\ref{fig:3}(d), and the red curve is the fitting.
    (c)~Calculated amplitude modulation $V_{\mathrm{d}}(t)$ and (d)~frequency modulation $\omega_{\mathrm{d}}(t)$ of the drive pulse to emit a photon with the desired shape and frequency.
    }
    \label{fig:4}
\end{figure}
As our target photon waveform, we choose a time-symmetric hyperbolic-secant function
\begin{equation}\label{target_shape}
    \psi_{\mathrm{\mathrm{target}}}(t)=\sqrt{\frac{\gamma_{\mathrm{ph}}}{2}}\,\mathrm{sech}(\gamma_{\mathrm{ph}} t),
\end{equation}
which has an exponentially decaying tail and is feasible to realize experimentally~\cite{kurpiers_deterministic_2018}.
The parameter $\gamma_{\mathrm{ph}}$ determines the temporal width of the generated photon.
Using Eq.~\eqref{gamma_t}, the corresponding time-dependent photon emission rate $\Gamma_f(t)$ takes the form
\begin{equation}\label{target_gamma}
    \Gamma_f^{\mathrm{target}}(t) = \frac{\gamma_{\mathrm{ph}}\mathrm{sech}^2(\gamma_{\mathrm{ph}}t)}{2(1-\tanh(\gamma_{\mathrm{ph}}t))}.
\end{equation}
Here, we describe the operational steps to design the drive pulse to emit a photon with the target shape $\psi_{\mathrm{target}}(t)$ and the target frequency of $\omega_{\mathrm{ph}}^{\mathrm{target}}/2\pi=10.32\mathrm{\,GHz}$. 

First, the dependence of $\Gamma_f$ on the drive amplitude $V_{\mathrm{d}}$ at the target frequency is extracted from the intersection points along the vertical line at $\omega_{\mathrm{ph}}^{\mathrm{target}}$ in Fig.~\ref{fig:3}(e), with the results plotted in Fig.~\ref{fig:4}(a).
These points are fitted to an even-polynomial function up to the tenth order of $V_\mathrm{d}$ (see Appendix~\ref{app:waveform_square_drive}).
Following a similar procedure, we determine the dependence of the drive frequency $\omega_{\mathrm{d}}$ on the drive amplitude $V_{\mathrm{d}}$ by identifying the intersection points along the horizontal line at $\omega_{\mathrm{ph}}^{\mathrm{target}}$ in Fig.~\ref{fig:3}(d), as shown in Fig.~\ref{fig:4}(b). 
The latter is also fitted with an even-polynomial function up to the tenth order.

To generate a photon with the target waveform, the drive amplitude $V_{\mathrm{d}}(t)$ needs to be modulated in the way which realizes the time-dependent photon emission rate $\Gamma^{\mathrm{target}}_f(t)$ given in Eq.~\eqref{target_gamma}. Such a modulation is given by $V_{\mathrm{d}}(t)=V_{\mathrm{d}}(\Gamma^{\mathrm{target}}_f(t))$, where the correspondence between $V_{\mathrm{d}}$ and $\Gamma_f$ can be obtained from Fig.~\ref{fig:4}(a).
As the drive amplitude $V_{\mathrm{d}}$ is modulated, the drive frequency $\omega_{\mathrm{d}}$ also needs to be modulated to compensate for the ac Stark shift. Such a modulation is given by $\omega_{\mathrm{d}}(t) = \omega_{\mathrm{d}}(V_{\mathrm{d}}(t))$, where the correspondence between $\omega_{\mathrm{d}}$ and $V_{\mathrm{d}}$ can be obtained from Fig.~\ref{fig:4}(b).
The resulting time-dependent drive parameters are shown in Figs.~\ref{fig:4}(c) and~(d). 
Applying the amplitude modulation $V_{\mathrm{d}}(t)$ and the frequency modulation $\omega_{\mathrm{d}}(t)$, the complex waveform of the drive pulse $\mathcal{V}_{\mathrm{d}}$ is given by
\begin{equation}\label{drive_pulse}
  \mathcal{V}_{\mathrm{d}}(t) =  V_{\mathrm{d}}(t) \exp\!\left(-i\int^t_0 \omega_{\mathrm{d}}(\tau)\,d\tau\right).
\end{equation}

The photon shape which can be generated is limited by the ability to realize specific values of $\Gamma_f$.
When emitting a photon in the sech-shaped mode [see Eq.~\eqref{target_shape}] at a frequency $\omega_{\mathrm{ph}}$, the width parameter $\gamma_{\mathrm{ph}}$ of the mode shape has to satisfy
\begin{equation}\label{maximum_gamma}
    \gamma_{\mathrm{ph}}\le \max_{V_{\mathrm{d}}} \frac{\Gamma_f(\omega_{\mathrm{ph}}, V_{\mathrm{d}})}{2}.
\end{equation}
According to Eq.~\eqref{Gamma_f_single}, the maximum $\Gamma_f$ which can be realized decreases as the target photon frequency deviates from the resonator frequency.
This gives rise to a trade-off relationship between the width parameter $\gamma_{\mathrm{ph}}$ and the frequency-tunable range.

\subsection{Emission and phase correction of shaped photons at different frequencies}

\begin{figure}[t]
    \centering
    \includegraphics{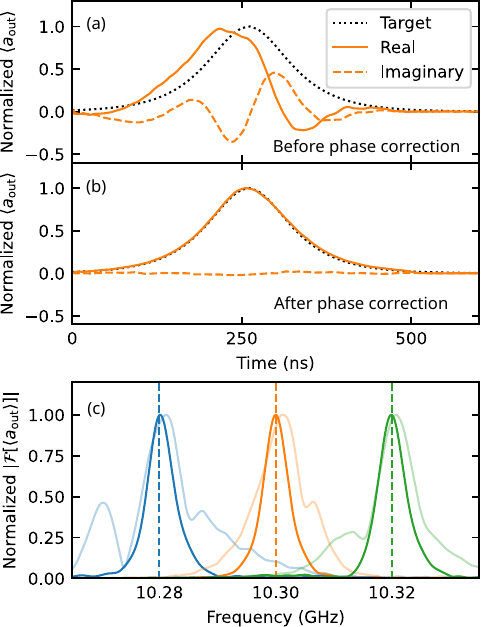}
    \caption{Waveforms and frequency components of the emitted shaped microwave photons. 
    The pulse sequence used here is the same as the one shown in Fig.~\ref{fig:3}(a) with the square drive pulse for the $\ket{f0}$--$\ket{g1}$ transition replaced with the shaped drive pulse in Figs.~\ref{fig:4}(c) and (d).
    (a), (b)~Measured photon shape at $\omega_{\mathrm{ph}}/2\pi=10.30$~GHz before and after the phase correction, respectively. The solid and dashed curves are the real and imaginary parts of the demodulated waveform, respectively. 
    Black dotted curve is the target photon shape $\psi_{\mathrm{ph}}^\mathrm{target}(t)$ given in Eq.~\eqref{target_shape}.
    (c)~Fourier amplitude of the photon emitted at different frequencies. The dashed lines indicate the corresponding target frequencies. Blue, orange and green curves correspond to the emitted photons with target frequencies of $\omega_{\mathrm{ph}}/2\pi=$10.28,~10.30,~and 10.32~GHz, respectively, after applying the phase correction. The pale curves correspond to the emitted photons before the phase correction.}
    \label{fig:5}
\end{figure}
To demonstrate the frequency tunability of our scheme while maintaining the target mode shape, we select target photon frequencies spanning a 40-MHz range around the resonator frequency: $\omega_{\mathrm{ph}}^{\mathrm{\mathrm{target}}}/2\pi\in  \{10.28,\,10.30,\,10.32\}$~GHz. We fix the width parameter to $\gamma_{\mathrm{ph}}/2\pi= 3\mathrm{\,MHz}$, which corresponds to a temporal width of approximately 500~ns.
Note that it is also possible to increase the tunable range by increasing the temporal width of the photon.

Figure~\ref{fig:5}(a) shows the result of the photon generation for the target frequency of $\omega_{\mathrm{ph}}^{\mathrm{target}}/2\pi=10.32\mathrm{\,GHz}$.
The real and imaginary parts of the photon waveform demodulated at the target frequency are plotted.
The observed phase distortion in the waveform is due to incomplete cancellation of the ac Stark shift at high drive amplitudes. 
We correct this phase distortion by applying a phase modulation to the control drive pulse based on the measured phase of the emitted photon.
We define the phase-corrected control drive $\mathcal{V}_{\mathrm{d}}^\mathrm{pc}(t)$ by modifying the original drive $\mathcal{V}_{\mathrm{d}}(t)$ as~\cite{yang_deterministic_2023}
\begin{equation}
    \mathcal{V}_{\mathrm{d}}^\mathrm{pc}(t) = \mathcal{V}_{\mathrm{d}}(t)\exp(-i\phi(t)).
\end{equation}
Here, $\phi(t)$ is the measured phase of the emitted photon under the original drive $\mathcal{V}_{\mathrm{d}}(t)$.
The phase-corrected photon waveform is shown in Fig.~\ref{fig:5}(b), where the near-zero imaginary part indicates a successful correction of the phase distortion. 
Figure~\ref{fig:5}(c) shows the Fourier amplitude of the acquired waveform at each of the target photon frequencies $\omega_{\mathrm{ph}}^{\mathrm{target}}/2\pi=\{10.28,\,10.30,\,10.32\}$~GHz.
After applying the phase correction, we observe that the photon frequencies are well aligned with the target frequencies, whereas the Fourier peaks are misaligned and distorted in the uncorrected case.

To quantify the time symmetry of the shaped photons, we calculate the metric $s$ defined as~\cite{pechal_microwave-controlled_2014}
\begin{equation}\label{time-sym}
    s = \max_{t_0} \left|\frac{\int|\ev{a_{\rm{out}}(t_0-t)}^*\ev{a_{\rm{out}}(t)}|\,dt}{\int|\ev{a_{\rm{out}}(t)}|^2\,dt}\right|,
\end{equation}
where $a_{\rm{out}}(t)$ is the output field from the resonator.
The values of $s$ calculated from the measured photon waveforms are shown in Table~\ref{tab1}.
Without the phase correction, the symmetries are below $84 \%$ for all the target frequencies.
The phase correction improves the symmetry to above 98\%, making the photon suitable for high-fidelity quantum communication.

\begin{table}[b]
    \caption{Time symmetry $s$ of the photon waveform.}
    \centering
    \begin{ruledtabular}
    \begin{tabular}{l|rrr}
    & \multicolumn{3}{c}{Target photon frequency} \\
    & $10.28\mathrm{\,GHz}$ &$10.30\mathrm{\,GHz}$&$10.32\mathrm{\,GHz}$ \\
    \hline
    Phase-uncorrected& $77.1(6)\%$&$81.2(5)\%$&$83.5(4)\%$ \\
    Phase-corrected&$98.4(2)\%$&$99.72(4)\%$&$99.0(1)\%$\\
    \end{tabular}
\end{ruledtabular}
    \label{tab1}
\end{table}

\subsection{Quantum state and process tomography}
\begin{figure}[t]
    \centering
    \includegraphics{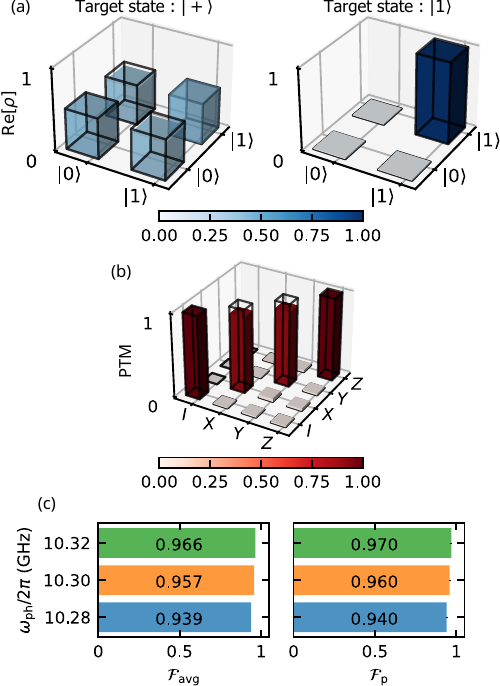}
    \caption{(a)~Reconstructed density matrices of the emitted photon mode at $\omega_{\mathrm{ph}}/2\pi=10.32$~GHz. Colored bars show the real parts of the reconstructed density matrices of the states $\ket{+}$~(left) and $\ket{1}$~(right) in the Fock basis.
    Black outlines show the ideal states.
    (b)~Measured Pauli transfer matrix~(PTM) of the quantum state transfer from the qubit to the emitted photon at 10.32~GHz. The red-colored bars represent the experimental result whereas the black outlines show the ideal matrix.
    (c)~Average state fidelities $\mathcal{F}_{\mathrm{avg}}$ of the six reconstructed density matrices of the emitted photons~(left) and the process fidelities $\mathcal{F}_{\mathrm{p}}$~(right) of the qubit-to-photon state transfer.}
    \label{fig:6}
\end{figure}
To confirm that the emitted photons are in the desired quantum states, we perform quantum state tomography of microwave photons emitted at each of the three target frequencies.
We apply a pump drive to the flux-driven JPA at twice the target frequency of the emitted photons to amplify one quadrature of the field.
The detection efficiency through the measurement chain, $\eta$, at each frequency is measured using a weak coherent pulse~\cite{kono_quantum_2018, sunada_efficient_2024}.
We obtain $\eta=0.374(1)$ for 10.28~GHz, $\eta=0.394(1)$ for 10.30~GHz, and $\eta=0.373(1)$ for 10.32~GHz.

To include waveform shaping accuracy in the tomography results, we use the target temporal mode $\psi_{\mathrm{\mathrm{target}}}(t)$ defined in Eq.~\eqref{target_shape}, rather than the actual wave packet measured in the experiment to extract the moments of the field. 
The annihilation operator $a$ in the chosen mode is defined as
\begin{equation} \label{pulse_mode}
a = \int \psi_{\mathrm{\mathrm{target}}}(t) a_{\rm{out}}(t) \, dt, 
\end{equation}
where $a_{\rm{out}}(t)$ is the output field operator.

We prepare the qubit in one of the six states $\ket{g}$, $(\ket{g} \pm \ket{f})/\sqrt{2}$, $(\ket{g} \pm i \ket{f})/\sqrt{2}$, and $\ket{f}$, and emit a photon respectively in the states $\ket{0}$, $\ket{\pm} = (\ket{0}\pm\ket{1})/\sqrt{2}$, $\ket{\pm i} = (\ket{0}\pm i\ket{1})/\sqrt{2}$, and $\ket{1}$, at each of the three target photon frequencies.
The density matrices of the emitted photons are reconstructed through quantum state tomography.
To verify the single-photon nature of the emitted radiation, we measure the fourth-order moments $\ev{a^\dag a^\dag aa}$ of the generated states~\cite{eichler_characterizing_2012}.
The average measured value is 0.003(11) (see Appendix~\ref{app:a4} for the individual values), confirming that the generated states are confined to the single-photon subspace.
Figure~\ref{fig:6}(a) shows the reconstructed density matrices for the states $\ket{+}$ and $\ket{1}$ at $\omega_{\mathrm{ph}}^{\mathrm{target}}/2\pi=10.32\mathrm{\,GHz}$.
We calculate the fidelities of the reconstructed density matrices to the target states as
\begin{equation} 
\mathcal{F} =\bra{\psi_{\mathrm{target}}}\rho\ket{\psi_{\mathrm{target}}}, 
\end{equation}
where $\rho$ is the reconstructed density matrix and $\ket{\psi_{\mathrm{target}}}$ is the target state. 
We calculate the average state fidelity $\mathcal{F}_{\mathrm{avg}}$ of the six states for each photon frequency, which is shown in the left panel of Fig.~\ref{fig:6}(c).

We also calculate the Pauli transfer matrix (PTM) to evaluate the process of the state transfer from the qubit to the emitted photon at each frequency.
The measured PTM at $\omega_{\mathrm{ph}}^{\mathrm{target}}/2\pi=10.32\mathrm{\,GHz}$ is shown in Fig~\ref{fig:6}(b).
The process fidelity $\mathcal{F}_{\mathrm{p}}$ is calculated as~\cite{chow_universal_2012}
\begin{equation}
    \mathcal{F}_{\mathrm{p}} =\frac{\mathrm{Tr}[R_{\mathrm{ideal}}^{\mathrm{T}}R] + d}{d + d^2},
\end{equation}
where $R$ is the estimated PTM,  $R_{\mathrm{ideal}} = I$ is the ideal PTM, which in this case is the identity matrix, and $d=4$ is the dimension of the Hilbert space under consideration.
Calculated process fidelities~$\mathcal{F}_{\mathrm{p}}$ are shown in the right panel of Fig.~\ref{fig:6}(c).

For both the average state fidelities $\mathcal{F}_{\mathrm{avg}}$ and the process fidelities $\mathcal{F}_{\mathrm{p}}$, we observe a small variation depending on the target photon frequency.
These may be due to the experimental instabilities of the room-temperature electronics and the thermal noise in the cable.
Nevertheless, the fidelities are near unity independently of the photon frequency, showing that a shaped photon can be accurately emitted regardless of the detuning between the photon frequency and the resonator frequency.

\section{Discussion}
We have demonstrated the generation of frequency-tunable, shaped single microwave photons using a fixed-frequency superconducting transmon qubit, achieving a tuning range of 40~MHz. 
This frequency tunability was realized through the implementation of the off-resonant driving scheme which stimulates a photon emission conditioned on the qubit state.
We characterized the frequency dependence of the qubit--transmission-line coupling via photon-emission spectroscopy performed with square drive pulses.
The spectroscopic results enabled us to design a drive pulse for generating a photon with a desired shape and frequency.
We compensated for the phase distortion of the emitted photon by modifying the drive pulse based on the phase of the measured photons.
The phase correction significantly improved the temporal symmetry of the photon wave packets, achieving symmetry metrics larger than 97\%.
Through quantum state and process tomography, we obtained consistently high average state fidelities and process fidelities around~95\% across the tunable range.
These results demonstrate the robustness of our approach. 

\begin{figure}[t]
    \centering
    \includegraphics{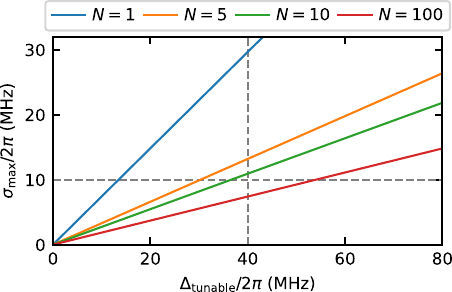}
    \caption{Maximum allowable resonator-frequency variation $\sigma_{\mathrm{max}}$ versus photon-frequency tunability $\Delta_{\mathrm{tunable}}$ that maintains successful quantum communication probability $P\geq0.5$ across $N$ fixed-frequency devices. 
    Different curves correspond to different number of devices ($N = 2,\,4,\,10,\,100,\,1000$). 
    The horizontal dashed line at $\sigma/2\pi = 10~\mathrm{MHz}$ indicates typical fabrication variations achievable with current technology. 
    The vertical dashed line displays the tuning range $\Delta_{\mathrm{tunable}}/2\pi=40~\mathrm{MHz}$ demonstrated in this work, which ensures reliable communication for networks up to $N = 10$ devices.
    }
    \label{fig:discussion}
\end{figure}

To evaluate the scalability of our approach, let us analyze the success probability $P$ of quantum communication across $N$ device pairs. 
We assume that fabrication uncertainty of the resonator frequency follows a Gaussian distribution with a standard deviation $\sigma$. 
Assuming that each device has a frequency tunability of $\Delta_{\mathrm{tunable}}$, the probability that all $N$ pairs can be frequency-matched is given by
\begin{equation}
    P = \mathrm{erf(\Delta_{\mathrm{tunable}}/2\sigma)}^N,
\end{equation}
where $\mathrm{erf}(x)$ is the error function.
Using this relationship, we determine the maximum tolerable fabrication variance $\sigma_{\mathrm{max}}$ to maintain $P\geq 0.5$ for different values of $N$ [Fig.~\ref{fig:discussion}].
Our demonstrated tuning range of $\Delta_{\mathrm{tunable}}/2\pi=40\,\mathrm{MHz}$ enables successful communication between up to $N=10$ pairs with $P\geq 0.5$, even with typical fabrication variations of $\sigma/2\pi = 10~\mathrm{MHz}$~\cite{valles-sanclemente_post-fabrication_2023, li_optimizing_2023}.
Scaling to larger networks requires improving the fabrication precision or increasing the frequency tunability.
In contrast, operating without the frequency tuning imposes severe constraints on the allowable frequency mismatch $\Delta\omega$ between the resonators.
For the temporal mode shape implemented in this work [Eq.~\eqref{target_shape}], achieving more than 99\% mode matching requires $\Delta\omega \lesssim 0.16\gamma_{\mathrm{ph}}$ (see Appendix~\ref{app_discussion}).
With $\gamma_{\mathrm{ph}}/2\pi = 3~\mathrm{MHz}$, this translates to a maximum allowable detuning of 0.48~MHz, far below the fabrication variations of $\sigma/2\pi \approx 10~\mathrm{MHz}$.
These estimates illustrate the advantage of our tunable scheme for implementing a hardware-efficient quantum communication network. 

Our characterization method based on the direct observation of the photon waveform establishes a direct mapping between the drive parameters and the emitted-photon parameters. This approach treats the system as a black box, removing the need for assumptions such as frequency-independent attenuation in input lines, perfect impedance matching, or strict adherence to idealized Hamiltonians. 
In the conventional photon-shaping methods, the communication fidelity is limited by the inaccuracies of these assumptions.
In contrast, our method can achieve improved shaping accuracy simply by refining the characterization measurements presented in Fig.~\ref{fig:3}, which captures the system's actual behavior under experimental conditions.

Whereas we have used one resonator mode to demonstrate our method, having multiple resonator modes simultaneously contribute to the photon-emission process can expand the capabilities of our scheme.
The collective contribution of multiple modes provides a broad spectral range for the resonator--transmission-line coupling, extending the tunable frequency range beyond our current demonstration.
The wide bandwidth can also accommodate multiple frequency channels, which are useful for more complex quantum networks which utilize frequency multiplexing.

The generation rate of the shaped photons warrants further discussion.
From Eq.~\eqref{maximum_gamma}, the photon-emission rate is limited by the maximally achievable $\Gamma_f(\omega_{\mathrm{ph}}, V_{\mathrm{d}})$ at the target photon frequency.
While $g_\mathrm{eff}$ and therefore $\Gamma_f$ can be increased to some extent with stronger driving, they are limited by the requirement $g_{\mathrm{eff}}\leq\kappa/4$ to avoid Rabi oscillations between the $\ket{f0}$ and $\ket{g1}$ states (see Appendix~\ref{app_rabif0g1}).
Although  our theoretical treatment in Sec.~\ref{chap2_idea} and Appendix~\ref{app_waveform} assumes $g_{\mathrm{eff}}\ll\kappa$, our experiments demonstrate stable operation near $g_{\mathrm{eff}}\approx\kappa/4$ (see Appendix~\ref{app:photon_shaping}).
This result suggests that, to accelerate photon emission, the drive strength can be increased up to the threshold $g_{\mathrm{eff}}=\kappa/4$ where Rabi oscillations begin to occur while still maintaining the effectiveness of our photon-shaping method.
However, a stronger drive may induce unwanted transitions and heating of the microwave environment. 
A comprehensive study of these trade-offs will be necessary for optimizing future implementations.

In our current implementation, a circulator enables direct observation of photon waveforms for calibrating the drive pulses. 
However, such circulators are a source of photon loss in communication experiments~\cite{kurpiers_deterministic_2018}. 
For future practical implementations, it will be beneficial to achieve accurate photon shaping without monitoring the photon waveform. 
While the measurement of the $\ket{f}$ state population under $\ket{f0}$--$\ket{g1}$ drive~\cite{magnard_fast_2018} can still be used to determine the dependence of the photon emission rate $\Gamma_f$ on drive amplitude $V_{\mathrm{d}}$ and drive frequency $\omega_{\mathrm{d}}$, characterizing the relationship between photon frequencies $\omega_{\mathrm{ph}}$ and the drive parameters ($V_{\mathrm{d}}$, $\omega_{\mathrm{d}}$) without direct detection remains challenging. 
Development of such measurement protocols which eliminate the need for circulators in communication schemes while maintaining high-fidelity photon shaping will be useful for practical quantum communication.

\begin{acknowledgments}
We thank S. Watanabe, K. Yuki, S. Inoue, S. Kikura and K. Sunada for fruitful discussions.
This work was supported in part by the University of Tokyo Forefront Physics and Mathematics Program to Drive Transformation (FoPM), a World-leading Innovative Graduate Study (WINGS) Program, the Ministry of Education, Culture, Sports, Science and Technology (MEXT) Quantum Leap Flagship Program (Q-LEAP) (Grant No.\ JPMXS0118068682), and the JSPS Grant-in-Aid for Scientific Research (KAKENHI) (Grant No.\ JP22H04937).
\end{acknowledgments}

\appendix

\section{Experimental setup}\label{app_setup}
\begin{figure}
    \centering
    \includegraphics{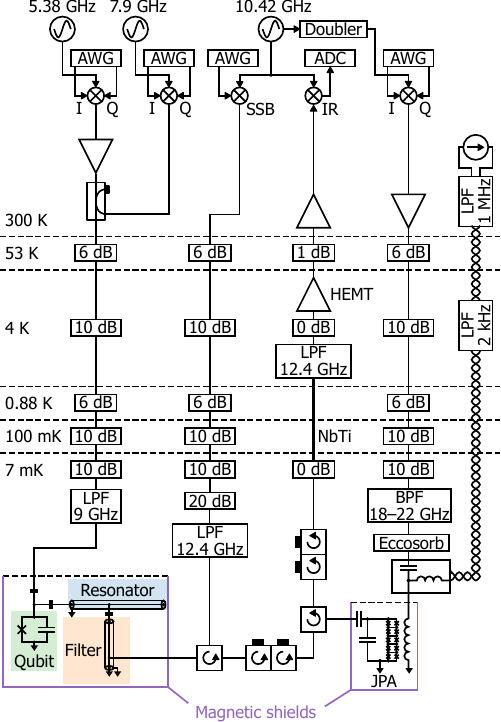}
    \caption{Measurement setup used in the experiment. AWG, arbitrary waveform generator; ADC, analog-to-digital converter; SSB, single sideband mixer; IR, image reject mixer; LPF, low-pass filter; HEMT, high-electron-mobility transistor; BPF, band-pass filter; JPA, Josephson parametric amplifier.}
    \label{fig:setup}
\end{figure}
\begin{table}[b]
    \caption{$T_1$ and $T_2$ of the qubit.}
    \centering
    \begin{ruledtabular}
    \begin{tabular}{lrrr}
    $\ket{g}$--$\ket{e}$ energy-relaxation time & $T_{1, ge}$ & $28.5\pm0.9\mathrm{\,\mu s}$\\
    $\ket{g}$--$\ket{e}$ Ramsey dephasing time & $T_{2, ge}^*$ & $5.7\pm0.3\mathrm{\,\mu s}$\\
    $\ket{g}$--$\ket{e}$ echo dephasing time & $T_{2, ge}^{\mathrm{e}}$ & $4.3\pm0.4\mathrm{\,\mu s}$\\
    $\ket{e}$--$\ket{f}$ energy-relaxation time & $T_{1,ef}$ & $23.0\pm1.5\mathrm{\,\mu s}$\\
    $\ket{e}$--$\ket{f}$ Ramsey dephasing time & $T_{2,ef}^*$ & $5.9\pm1.4\mathrm{\,\mu s}$\\
    \end{tabular}
    \end{ruledtabular}
    \label{tab_para_qubit}
\end{table}
\begin{table}[tbh]
    \caption{Parameters of the bare resonator and filter.}
    \centering
    \begin{ruledtabular}
    \begin{tabular}{lrrr}
    Resonator frequency & $\omega_{\mathrm{r},g}^{\mathrm{b}}/2\pi$ & $10.287\mathrm{\,GHz}$\\
    Filter frequency & $\omega_{\mathrm{f}}^{\mathrm{b}}/2\pi$ & $10.179\mathrm{\,GHz}$\\
    Coupling strength & $J/2\pi$ & $84.1\mathrm{\,MHz}$\\
    External coupling of the filter & $\kappa_{\mathrm{f}}/2\pi$ & $400\mathrm{\,MHz}$\\
    Qubit--resonator full dispersive shift & $\chi/2\pi$ & $-4.5\mathrm{\,MHz}$\\
    \end{tabular}        
    \end{ruledtabular}
    \label{tab_para}
\end{table}

The setup of this work is shown in Fig~\ref{fig:setup}.
The frequency of local oscillators are set so that they satisfy the equation~\cite{ilves_-demand_2020}
\begin{equation}
    \omega_{f0g1}^{\mathrm{LO}} + \omega_{\mathrm{r}}^{\mathrm{LO}}  - 2\omega_{\mathrm{q}}^{\mathrm{LO}} = 0,
\end{equation}
where $\omega_{f0g1}^{\mathrm{LO}}/2\pi=5.380\mathrm{\,GHz}$, $\omega_{\mathrm{r}}^{\mathrm{LO}}/2\pi=10.420\mathrm{\,GHz}$, and 
$\omega_{\mathrm{q}}^{\mathrm{LO}}/2\pi=7.900\mathrm{\,GHz}$ are the frequencies of the local oscillators for the $\ket{f0}$--$\ket{g1}$ drive, down conversion and qubit drive, respectively.
The drive line used for the photon emission is merged with the qubit control line at room temperature by a directional coupler.

The coherence times of the qubit are listed in Table~\ref{tab_para_qubit}.
The bare parameters of the resonator and the filter corresponding to the fitting result in Fig.~\ref{fig:2}(c) are listed in Table~\ref{tab_para}.
We fit the data with $S_{11,g}/S_{11,e}$, where
\begin{equation}
    S_{11, g(e)} = \cos\theta+e^{i\theta}\frac{i\kappa_{\mathrm{f}}\left(\omega-\omega_{\mathrm{r},g(e)}^{\mathrm{b}}\right)}{((\omega-\omega_{\mathrm{f}}^{\mathrm{b}})-i\kappa_{\mathrm{f}}/2)(\omega-\omega_{\mathrm{r},g(e)}^{\mathrm{b}})+J^2}.
\end{equation}
Here, $\theta$ accounts for the impedance mismatch in the measurement line~\cite{swiadek_enhancing_2024}. 
The superscript `$\mathrm{b}$' represents the bare parameter, and $\omega_{\mathrm{r},e}^{\mathrm{b}}=\omega_{\mathrm{r},g}^{\mathrm{b}}+\chi$.
We can calculate the frequencies and the linewidths of hybridized modes using the equations below~\cite{swiadek_enhancing_2024}:
\begin{align}
    \omega_{\pm} &= \frac{\omega_{\mathrm{r}}^{\mathrm{b}} + \omega_{\mathrm{f}}^{\mathrm{b}}}{2} \pm \frac{1}{2} \mathrm{Re}\sqrt{\Xi}, \label{A3} \\
    \kappa_{\pm} &= \frac{\kappa_{\mathrm{f}}}{2} \mp \mathrm{Im}\sqrt{\Xi}.\label{A4}
\end{align} 
Here, we introduce $\Xi = ((\omega_{\mathrm{r}}^{\mathrm{b}}-\omega_{\mathrm{f}}^{\mathrm{b}})+i\kappa_{\mathrm{f}}/2)^2+4J^2$.
By substituting the resonator paramters in Eqs.~\eqref{A3} and \eqref{A4}, we obtain the characteristic parameters for the two hybridized modes as $(\omega_{+}/2\pi,\,\kappa_{+}/2\pi) = (10.306\mathrm{\,GHz},\,53.3\mathrm{\,MHz})$ and $(\omega_{-}/2\pi,\,\kappa_{-}/2\pi) = (10.160\mathrm{\,GHz},\,347\mathrm{\,MHz})$.
We neglect the mode with $(\omega_{-}/2\pi,\,\kappa_{-}/2\pi)$ since its linewidth is much larger and has much faster dynamics than the other one.

\section{Photon-emission process}\label{app_waveform}
To derive the waveforms of emitted photons, we introduce the master equation of the system.
The Hamiltonian of the system is written as follows:
\begin{equation}
\begin{split}
    \mathcal{H}/\hbar = &\omega_{\mathrm{r}} c^\dag c + \omega_{ge} b^\dag b + \frac{\alpha}{2} b^\dag b^\dag b b + g(c^\dag b + c b^\dag) \\
    &+ \Omega_{\mathrm{d}}(t)\cos(\omega_{\mathrm{d}} t) (b + b^\dag)\\
    &+\sqrt{\frac{\kappa}{2\pi}}\int (c^\dag a_\omega +c a_\omega^\dag)\,d\omega + \int \omega a_\omega^\dag a_\omega\,d\omega,
\end{split}
\end{equation}
where $c$, $b$, and $a_\omega$ are the annhilation operators of a resonator, a qubit, and a transmission-line mode at frequency of $\omega$.
We move to a rotating frame at $\omega_{\mathrm{d}}$ and perform the Schrieffer-Wolff transformation. 
Considering the subspace composed of three states $\{\ket{f0},\,\ket{g1},\,\ket{g0}\}$, we obtain~\cite{pechal_microwave-controlled_2014}
\begin{equation}
\begin{split}
    \mathcal{H}/\hbar = \delta_{f0}\ket{f0}\!\bra{f0} + \delta_{g1}\ket{g1}\!\bra{g1} \\
    + g_{\mathrm{eff}}(\ket{f0}\!\bra{g1} + \mathrm{h.c.}).
\end{split}
\end{equation}
Here, $\delta_{f0}=2\omega_{ge}+\alpha-2\omega_{\mathrm{d}}$, $\delta_{g1}=\omega_{\mathrm{r}}-\omega_{\mathrm{d}}$, and
\begin{equation}
    g_{\mathrm{eff}} = \frac{g\alpha\Omega_{\mathrm{d}}(t)}{\sqrt{2}(\omega_{\mathrm{r}}-\omega_{ge})(\omega_{\mathrm{r}}-\omega_{ge}-\alpha)}.
\end{equation}
We also omit the coupling term between $\ket{g0}$ and $\ket{g1}$ given that the drive frequency is far detuned from this transition frequency.
In the following, we take $g_{\mathrm{eff}}$ as real-valued.

We calculate the dynamics of the system by solving the Lindblad master equation:
\begin{equation}\label{master_eq}
    \dot{\rho} = i[\mathcal{H}/\hbar,\, \rho] + \kappa\mathcal{D}_{\rho}[c],
\end{equation}
where $\mathcal{D}$ is the superoperator that governs the resonator decay:
\begin{equation}
    \mathcal{D}_{\rho}[c] = c\rho c^\dag - \frac{1}{2}\{c^\dag c,\,\rho\}.
\end{equation}
We can rewrite the Hamiltonian and the density matrix as follows:
\begin{equation}
    \mathcal{H}/\hbar = \begin{pmatrix}
        \delta_{f0} & g_{\mathrm{eff}} & 0 \\
        g_{\mathrm{eff}} & \delta_{g1} & 0 \\
        0 & 0& 0
    \end{pmatrix},
\end{equation}
\begin{equation}
    \rho = \begin{pmatrix}
        \rho_{{f0,f0}} & \rho_{{f0,g1}} & \rho_{{f0,g0}} \\
        \rho_{{g1,f0}} & \rho_{{g1,g1}} & \rho_{{g1,g0}} \\
        \rho_{{g0,f0}} & \rho_{{g0,g1}} & \rho_{{g0,g0}}
    \end{pmatrix}.
\end{equation}
Using this notation, we rewrite Eq.~\eqref{master_eq} as
\begin{subequations}
\begin{align}
    \dot{\rho}_{{f0,f0}} &= -2\mathrm{Im}[g_{\mathrm{eff}}\,\rho_{{f0,g1}}] \label{f0f0}, \\
    \dot{\rho}_{{g1,g1}} &= 2\mathrm{Im}[g_{\mathrm{eff}}\,\rho_{{f0,g1}}] - \kappa\rho_{{g1,g1}}\label{g1g1}, \\
    \dot{\rho}_{{g0,g0}} &= \kappa \rho_{{g1,g1}}\label{f0g1}, \\
    \dot{\rho}_{{f0,g1}} &= -i(\Delta_{\mathrm{d}}\,\rho_{{f0,g1}} + g_{\mathrm{eff}}(\rho_{{g1,g1}}-\rho_{{f0,f0}}))-\frac{\kappa}{2}\rho_{{f0,g1}}, \\
    \dot{\rho}_{{f0,g0}} &= -i(\delta_{{f0}}\rho_{{f0,g0}}+g_{\mathrm{eff}}\rho_{{g1,g0}}) \label{f0g0},\\
    \dot{\rho}_{{g1,g0}} &= -i(\delta_{{g1}}\rho_{{g1,g0}}+g_{\mathrm{eff}}\rho_{{f0,g0}})-\frac{\kappa}{2}\rho_{{g1,g0}}\label{g1g0},
\end{align}
\end{subequations}
where $\Delta_{\mathrm{d}}=\delta_{f0}-\delta_{g1}=\omega_{f0g1}-\omega_{\mathrm{d}}$ is the difference between the drive frequency and the $\ket{f0}$--$\ket{g1}$ transition frequency.

\subsection{Maximum coupling strength between ${\boldsymbol{\ket{f0}}}$ and $\boldsymbol{\ket{g1}}$}\label{app_rabif0g1}
To ensure that our photon-shaping method works, we need to prevent Rabi oscillations between the states $\ket{f0}$ and $\ket{g1}$ under the applied drive.
Here, we derive the condition on $g_{\mathrm{eff}}$ that suppresses such oscillations.
For simplicity, we consider the case where $\Delta_{\mathrm{d}}=0$ and denote $\mathrm{Im}[\rho_{f0,g1}(t)]$ as $u(t)$.
From Eqs.\eqref{f0f0}, \eqref{g1g1}, and \eqref{f0g1}, we obtain
\begin{equation}
    \frac{d}{dt} \begin{pmatrix}
        \rho_{f0,f0} \\ \rho_{g1,g1} \\ u
    \end{pmatrix}
    =
    \begin{pmatrix}
        0 & 0 & -2g_{\mathrm{eff}} \\
        0 & -\kappa & 2g_{\mathrm{eff}} \\
        g_{\mathrm{eff}} & -g_{\mathrm{eff}} & -\kappa/2
    \end{pmatrix}
    \begin{pmatrix}
        \rho_{f0,f0} \\ \rho_{g1,g1} \\ u
    \end{pmatrix}.
\end{equation}
The eigenvalues $\lambda_1$, $\lambda_+$, and $\lambda_-$ of the matrix on the right hand side characterize the temporal evolution.
\begin{subequations}
    \begin{align}
        \lambda_1 &= -\frac{\kappa}{2},\\
        \lambda_{\pm} &=  \frac{-\kappa \pm \sqrt{\kappa^2 - 16g_{\mathrm{eff}}^2}}{2}.
    \end{align}
\end{subequations}
When $g_{\mathrm{eff}} \leq \kappa/4$, all eigenvalues are real, indicating overdamped dynamics.
On the other hand, when $g_{\mathrm{eff}} > \kappa/4$, the eigenvalues $\lambda_{\pm}$ acquire imaginary components which indicate the onset of Rabi oscillations.
Therefore, to suppress Rabi oscillations between $\ket{f0}$ and $\ket{g1}$, we need to satisfy 
\begin{equation}
    g_{\mathrm{eff}} \leq \kappa/4
\end{equation}
This inequality establishes an upper bound on the drive strength for a proper photon emission.

\subsection{Photon waveforms under square drives}\label{app:waveform_square_drive}
The waveform of an emitted photon can be directly acquired using the input--output relation \cite{gardiner_input_1985}
\begin{equation}
    a_{\mathrm{out}} = a_{\mathrm{in}} + \sqrt{\kappa}\,c.
\end{equation}
Here, $a_{\mathrm{in}}$ is the input field operator to the resonator, and $a_{\mathrm{out}}$ is the output operator from the resonator. 
By taking the expectation values of both sides of the equation and assuming that the input is in a vacuum state, i.e., $\ev{a_{\mathrm{in}}}=0$, we find $\ev{a_{\mathrm{out}}(t)} = \sqrt{\kappa}\,\ev{a(t)} = \sqrt{\kappa}\,\rho_{{g1,g0}}$.
Therefore, we obtain the waveform of the emitted photon by solving the equations for $\rho_{g1,g0}$

Since Eqs.~\eqref{f0g0} and \eqref{g1g0} are independent of the other differential equations, we can get an analytic solution for $\rho_{{g1,g0}}$ using these equations when $g_{\mathrm{eff}}$ is not dependent on time:
\begin{equation}\label{square_waveform}
   \rho_{{g1,g0}} = C_1 \exp[-(\gamma_-+i\omega_-)t] + C_2\exp[-(\gamma_+ + i\omega_+)t],
\end{equation}
where $C_1$ and $C_2$ are the coefficients determined by the initial state of the system.
Here, we define the decay rates and the frequencies of the photon,
\begin{align}
    \gamma_{\pm} &= \frac{\kappa}{4}\pm(X^2+Y^2)^{\frac{1}{4}}\cos(\theta/2), \\
    \omega_{\pm} &= \frac{\delta_{{f0}}+\delta_{{g1}}}{2}
    \pm(X^2+Y^2)^{\frac{1}{4}}\sin(\theta/2),
\end{align}
where $X=\frac{\kappa^2}{16}-g_{\mathrm{eff}}^2-\frac{\Delta_{\mathrm{d}}^2}{4}$, $Y=\frac{\kappa\Delta_{\mathrm{d}}}{4}$ and $\theta=\mathrm{arctan}(Y/X)$.
According to Eq.~\eqref{square_waveform}, the emitted photon has basically two contributions to its waveform.
However, in the case of  $\kappa \gg g_{\mathrm{eff}}$, $\exp(-\gamma_+ t)$ decays significantly faster than $\exp(-\gamma_- t)$.
This disparity allows us to approximate the emitted waveform as an exponentially decaying shape with a decay rate of $\gamma_-$.
When expanding $\gamma_-$ and $\omega_-$ with respect to $g_{\mathrm{eff}}$, there are only even-order terms, which is the reason of fitting with an even  polynomial function in Fig.~\ref{fig:4}.

\subsection{Temporal photon shaping}\label{app:photon_shaping}
We assume that the resonator dynamics can be neglected (\(\dot{\rho}_{{g1,g1}} = \dot{\rho}_{{f0,g1}} = 0\)) when the linewidth of the resonator $\kappa$ is sufficiently large.
In this case, we obtain
\begin{subequations}
\begin{align}
    \dot{\rho}_{f0,f0}(t) &= -\Gamma_f(t) (\rho_{f0,f0}(t)-\rho_{g1,g1}), \label{d_f0f0_simp}\\
    \rho_{g1,g1} &= \frac{2g_{\mathrm{eff}}}{\kappa}\mathrm{Im}[\rho_{f0, g1}], \label{g1}
\end{align}
\end{subequations}
where
\begin{equation}\label{Gamma_f}
\Gamma_f = \frac{\kappa g_{\mathrm{eff}}^2}{(\omega_{f0g1}-\omega_{\mathrm{d}})^2 + (\kappa/2)^2}.
\end{equation}
Because we have $\omega_{f0g1}=\omega_{ge}+\omega_{ef}-\omega_{\mathrm{r}}$, we arrive at Eq.~\eqref{Gamma_f_single}.
Since $\rho_{f0,f0}$ and $\rho_{g1,g1}$ are the populations of $\ket{f0}$ and $\ket{g1}$, respectively, we have the excitation conservation relation:
\begin{equation}\label{energy_conserve}
    \rho_{f0,f0}(t) = 1 - \int^t dt^\prime |\psi_{\mathrm{ph}}(t^\prime)|^2 - \rho_{g1,g1},
\end{equation}
where $\psi_{\mathrm{ph}}(t)$ is the temporal mode function of the emitted photon.
When $\kappa\gg g_{\mathrm{eff}}$, the population of $\ket{g1}$ is close to zero, i.e., $\rho_{g1,g1}\approx 0$.
Thus, $\Gamma_f$ is considered as the decay rate of the $\ket{f0}$ state under the drive pulse as Eq.~\eqref{d_f0f0_simp} shows.
With this condition, we derive Eq.~\eqref{gamma_t} using Eqs.~\eqref{d_f0f0_simp} and \eqref{energy_conserve}.
Furthermore, with $\kappa\gg g_{\mathrm{eff}}$, we obtain \(\gamma_- \simeq \Gamma_f / 2\). 
This relationship allows us to estimate the decay rate \(\Gamma_f\).

In the arguments above, we have assumed the condition for the adiabatic elimination of the resonator $\kappa\gg g_{\mathrm{eff}}$ to make our method work.
This condition is required to effectively shape the temporal profile of the photon emission.
In our experiments, we used the drive amplitudes up to 
$V_{\mathrm{d}}=1.2\mathrm{\,V}$, which corresponds to the drive strength for the $\ket{g}$--$\ket{e}$ transition of $\Omega_{\mathrm{d}}/2\pi\approx1.7$~GHz and the effective coupling strength of $g_{\mathrm{eff}}/2\pi\approx 13~$MHz.
The effective coupling strength was determined through the ac Stark shift measurements~\cite{zeytinoglu_microwave-induced_2015}.
This $g_{\mathrm{eff}}$ value is close to the threshold $g_{\mathrm{eff}}=\kappa/4$, beyond which Rabi oscillations between $\ket{f0}$ and $\ket{g1}$ take place.
Regardless of this strong drive, we have confirmed that our method still works.

\section{Quantum state tomography of a single microwave photon}\label{app_QST}
\begin{table}[t]
    \caption{$\ev{a^\dag a^\dag a a}$ for the photons without phase correction.}
    \begin{ruledtabular}
    \begin{tabular}{lrrr}
    State & 10.28~GHz & 10.30~GHz & 10.32~GHz \\
    \hline
    $\ket{+}$ & $-0.007(14)$& 0.021(10)& 0.006(13) \\
    $\ket{-}$ & 0.001(14)& 0.023(10)&  $-0.002(13)$  \\
    $\ket{+i}$ & $-0.013(14)$& 0.021(10)&  0.002(13)  \\
    $\ket{-i}$ & 0.006(14)& 0.022(10)&  $-0.001(13)$  \\
    $\ket{1}$ & 0.025(17)& 0.027(12)&  $-0.002(15)$  \\
    $\ket{0}$ & 0.009(11)& 0.013(7)&    0.001(10)
    \end{tabular}
\end{ruledtabular}
    \label{tab_a4}
\end{table}

\begin{table}[t]
    \caption{$\ev{a^\dag a^\dag a a}$ for the photons with phase correction.}
    \centering
    \begin{ruledtabular}
    \begin{tabular}{lrrr}
    State & 10.28~GHz & 10.30~GHz & 10.32~GHz \\
    \hline
    $\ket{+}$ & 0.000(11)& 0.009(10)& 0.000(12) \\
    $\ket{-}$ & 0.015(11)& 0.005(10)&  0.004(12)  \\
    $\ket{+i}$ & $-0.007(11)$& 0.004(10)&  0.002(12)  \\
    $\ket{-i}$ & 0.007(11)& 0.004(10)&  $-0.009(11)$  \\
    $\ket{1}$ & 0.008(13)& 0.007(12)&  $-0.008(14)$  \\
    $\ket{0}$ & 0.007(8)& 0.009(7)&    0.002(8)
    \end{tabular}
\end{ruledtabular}
    \label{tab_a4u}
\end{table}

\begin{figure}[t]
    \centering
    \includegraphics{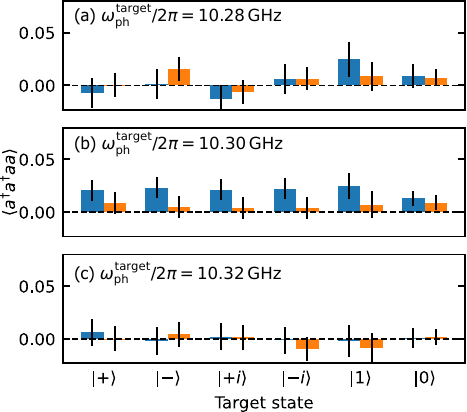}
    \caption{Fourth-order moments $\ev{a^\dag a^\dag aa}$ of the emitted photon at the target photon frequency $\omega_{\mathrm{ph}}^{\mathrm{target}}/2\pi$ of (a)~10.28~GHz, (b)~10.30~GHz, and (c)~10.32~GHz. 
    Blue bars are for the photons without phase correction, and orange bars are for the photons with phase correction.}
    \label{fig_a4}
\end{figure}

We perform quantum state tomography of the emitted photons through the quadrature measurements.
Here, we define the field quadrature operators as
\begin{equation}
    q_{\theta} = \frac{a e^{-i\theta} + a^\dag e^{i\theta}}{\sqrt{2}}.
\end{equation}
We follow the method in Ref.~\citenum{sunada_efficient_2024} to reconstruct the density matrices.
In this method, we assume the density operator of an emitted photon to be
\begin{equation}
    \rho = \frac{1}{2}(I + c_x X + c_y Y + c_z Z).
\end{equation}
Here, $c_x,\,c_y,\,c_z$ are the real-valued coefficients and $I,\,X,\,Y,\,Z$ are the Pauli operators defined as
\begin{subequations}
    \begin{align}
        I &= \ket{0}\!\bra{0} + \ket{1}\!\bra{1}, \\
        X &= \ket{1}\!\bra{0} + \ket{0}\!\bra{1}, \\
        Y &= i\ket{1}\!\bra{0} - i\ket{0}\!\bra{1}, \\
        Z &= \ket{0}\!\bra{0} - \ket{1}\!\bra{1}.
    \end{align}
\end{subequations}
We reconstruct the density matrix by obtaining the coefficients following the equations.
\begin{subequations}
    \begin{align}
        c_x &= \ev{X} = \sqrt{2}\,\ev{q_{\theta=0}}, \\
        c_y &= \ev{Y} = \sqrt{2}\,\ev{q_{\theta=\pi/2}}, \\
        c_z &= \ev{Z} = 2-\ev{q_{\theta=0}^2}-\ev{q_{\theta=\pi/2}^2}.
    \end{align}
\end{subequations}

We note that our density matrix reconstruction method does not impose any constraints that would ensure the resulting matrix is physically valid (i.e., positive semidefinite). We directly calculate the density matrix elements from measured expectation values of quadrature operators, allowing the experimental noise and uncertainties to be reflected in the results. By not imposing such constraints, our method offers a more transparent view of the raw experimental data, but it means the reconstructed density matrices may occasionally have unphysical properties such as fidelities exceeding unity.

\subsection{Stabilizing JPA gain via PID control}
When performing quantum state tomography, we operate the JPA in a high-gain regime, where its gain is highly sensitive to the pump drive strength. 
Under the experimental conditions, pump-strength fluctuations occur due to instrumental instabilities and environmental factors such as laboratory-temperature variations from the air conditioning systems. 
These fluctuations make it crucial to maintain a stable JPA gain during quadrature measurements.
To stabilize the gain, we implement a PID feedback control system that monitors quadrature measurements of the vacuum state, which we also use for normalizing the photon field operator. 
The vacuum-state measurements results in a distribution in the IQ plane, where the standard deviation along the amplification direction corresponds to the JPA gain, and the orientation of the distribution indicates the amplification axis. 
The PID controller adjusts both the phase and amplitude of the JPA pump signal to maintain the desired gain and amplification direction by continuously tracking these distribution parameters.

\begin{table}[t]
    \centering
    \caption{Fidelities of the photon states without phase correction.}
    \centering
    \begin{ruledtabular}
    \begin{tabular}{lrrr}
    State & 10.28~GHz & 10.30~GHz & 10.32~GHz \\
    \hline
    $\ket{+}$ & 0.8248(9)& 0.8843(9)& 0.8959(9) \\
    $\ket{-}$ & 0.8230(9)& 0.8818(9)& 0.8985(9)  \\
    $\ket{+i}$ & 0.8265(9)& 0.8705(9)& 0.8929(9)  \\
    $\ket{-i}$ & 0.8250(9)& 0.8705(9)& 0.8865(9)  \\
    $\ket{1}$ & 0.536(2)& 0.700(2)& 0.763(2)  \\
    $\ket{0}$ & 1.000(2)& 1.001(1)&  1.000(1)
    \end{tabular}
\end{ruledtabular}
    \label{tab_fu}
\end{table}
\begin{table}[t]
    \caption{Fidelities of the photon states with phase correction.}
    \centering
    \begin{ruledtabular}
    \begin{tabular}{lrrr}
    State & 10.28~GHz & 10.30~GHz & 10.32~GHz \\
    \hline
    $\ket{+}$ & 0.9264(9)& 0.9432(8)& 0.9541(9) \\
    $\ket{-}$ & 0.9275(9)& 0.9453(8)&  0.9507(9)  \\
    $\ket{+i}$ & 0.9303(9)& 0.9483(8)&  0.9467(9)  \\
    $\ket{-i}$ & 0.9273(9)& 0.9392(8)&  0.9507(9)  \\
    $\ket{1}$ & 0.918(2)& 0.962(2)&  0.990(2)  \\
    $\ket{0}$ & 1.002(1)& 1.004(1)&    1.003(1)
    \end{tabular}
\end{ruledtabular}
    \label{tab_f}
\end{table}

\subsection{Single-photon validation}\label{app:a4}

To validate that the emitted microwave photons do not contain more than one photons, we estimate the fourth-order moment $\ev{a^\dag a^\dag aa}$.
Then, the probability that the pulse contains two or more photons can be upper-bounded as
\begin{equation}
    \sum_{n=2}^\infty \bra{n}\rho\ket{n} \le \frac{1}{2}\ev{a^\dag a^\dag aa}.
\end{equation}
We calculate the fourth-order moment using the moments of the measured quadratures as
\begin{equation}
    \ev{a^\dag a^\dag aa} = \frac{1}{2} + \sum_{\theta = 0,\,\frac{\pi}{4},\,\frac{\pi}{2},\,\frac{3\pi}{4}}\left(\frac{1}{6}\ev{q_{\theta}^4} - \frac{1}{2}\ev{q_{\theta}^2}\right).
\end{equation}
To take into account the measurement inefficiency, we calculate
\begin{equation}
    \ev{a^\dag a^\dag aa} = \eta^2\!\ev{a^\dag a^\dag aa}_{\mathrm{meas}}.
\end{equation}
The measured moments are shown in Fig.~\ref{fig_a4} and listed in Tables~\ref{tab_a4} and \ref{tab_a4u}.
The results of the phase-corrected photons are taken on a different day from those of the phase-uncorrected photons.
For photons at $\omega_{\mathrm{ph}}/2\pi=10.30~\mathrm{GHz}$ without phase correction, we observe systematically higher values compared to other results, which may be attributed to the back-action of the JPA pump~\cite{kindel_generation_2016} and/or the thermal noise in the cable.
Nevertheless, they are close to zero regardless of the states or the frequencies, supporting that the emitted pulses are in the $\{\ket{0},\,\ket{1}\}$ subspace. 

\begin{figure}[t]
    \centering
    \includegraphics{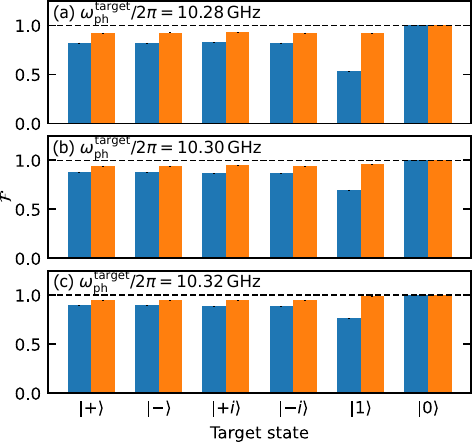}
    \caption{State fidelity in each target state at the photon frequency of (a)~10.28~GHz, (b)~10.30~GHz, and (c)~10.32~GHz.
    The blue bars are for the photons without phase correction, while the orange bars shows the result for phase-corrected photons.}
    \label{fig_QST_all}
\end{figure}

\begin{figure}[t]
    \centering
    \includegraphics{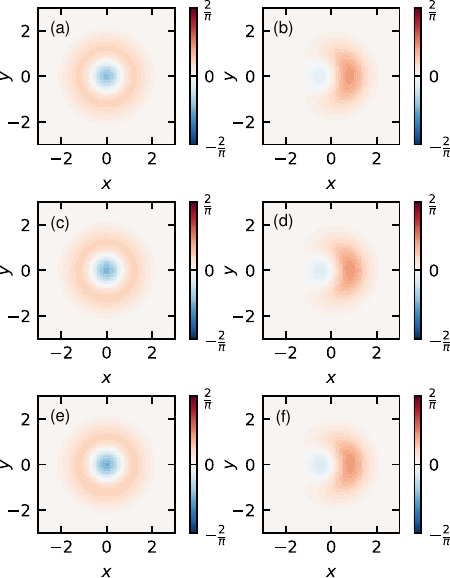}
    \caption{Wigner functions of the reconstructed photonic states for (a)(c)(e)~$\ket{1}$ and (b)(d)(f)~$\ket{+}$ at the frequency of (a)(b)~10.28~GHz, (c)(d)~10.30~GHz, and (e)(f)~10.32~GHz.}
    \label{fig_wigner}
\end{figure}

\subsection{Results of quantum state tomography for each state and frequency}
We present additional tomography results that complement the main text. Figure~\ref{fig_QST_all} displays the state fidelity for each target state and frequency, with the values listed in Tables~\ref{tab_fu} and~\ref{tab_f}. The fidelities for phase-uncorrected photons are systematically lower than those with phase correction, demonstrating the effectiveness of our phase correction protocol. 
We observe that the ground-state fidelities slightly exceed unity, which can happen if the measured moments $\ev{q_{\theta=0}^2},\,\ev{q_{\theta=\pi/2}^2}$ are unexpectedly small. 
This result is a consequence of our reconstruction method not imposing positive-semidefinite constraints on the density matrix. 
When experimental uncertainties such as thermal noise and temporal drift of the pump signals affect the measurements, the resulting density matrix can have non-physical properties. 
We believe presenting the results without additional mathematical constraints provides a more accurate reflection of the experimental performance.
Despite these phenomena, each state fidelity exceeds 91\%, validating the robustness of our photon generation method.
We also plot the Wigner functions of the states $\ket{1}$ and $\ket{+}$ at each target frequency in Fig.~\ref{fig_wigner}.

\section{Impact of the frequency difference between two devices on quantum communication}\label{app_discussion}

\begin{figure}[h]
    \centering
    \includegraphics{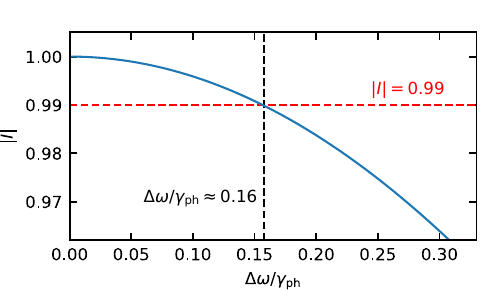}
    \caption{Mode-matching efficiency between the sender and the receiver as a function of the normalized frequency difference $\Delta\omega/\gamma_{\mathrm{ph}}$ for fixed-frequency devices. 
    The overlap integral $I$ quantifies the absorption efficiency at the receiver. 
    The red dashed line indicates $|I| = 0.99$, allowing high-fidelity quantum communication. 
    The vertical dashed line is the threshold at $\Delta\omega/\gamma_{\mathrm{ph}}\approx0.16$ to achieve $|I| = 0.99$.
    }
    \label{fig_overlap}
\end{figure}

Here, we analyze the maximum permissible frequency mismatch between fixed-frequency devices for successful quantum communication without frequency tuning capabilities of emitted photons. 
We calculate the mode overlap $I$ between the emitted photon from the sender~(frequency $\omega_1$) and the absorption mode of the receiver~(frequency $\omega_2$).

\begin{equation}
    I = \int_{-\infty}^\infty |\psi_{\mathrm{target}}(t)|^2 \exp(-i\Delta\omega t)\,dt,
\end{equation}
where $\Delta\omega=\omega_1-\omega_2$ is the frequency difference between the devices and $\psi_{\mathrm{target}}(t)$ is the temporal shape of photons~[Eq.~\eqref{target_shape}].
The overlap function $I$ quantifies the achievable absorption efficiency at the receiver.
Our numerical analysis reveals that maintaining $|I|\geq0.99$ requires $\Delta\omega/\gamma_{\mathrm{ph}}\leq0.16$ (Fig.~\ref{fig_overlap}).
For our experimental parameter, $\gamma_{\mathrm{ph}}/2\pi = 3~\mathrm{MHz}$, the condition translates to a maximum allowable frequency mismatch of approximately 0.48~MHz, which is significantly smaller than the typical fabrication variations of $\sigma/2\pi\approx 10~\mathrm{MHz}$.

\FloatBarrier
\nocite{*}
\bibliography{TunablePhoton}  

\end{document}